\documentclass[aps,prf,amsmath,amssymb,longbibliography,superscriptaddress]{revtex4-1}
\usepackage[dvipsnames,rgb,dvips]{xcolor}
\usepackage[dvipsnames,rgb,dvips]{xcolor}
\usepackage{overpic}
\usepackage{graphicx}
\usepackage{amssymb}
\usepackage{dcolumn}
\usepackage{epstopdf}
\usepackage{siunitx}
\usepackage{mathrsfs}
\makeatletter
\def\amsbb{\use@mathgroup \M@U \symAMSb}
\makeatother
\usepackage{bbm}
\expandafter\let\csname equation*\endcsname\relax
\expandafter\let\csname endequation*\endcsname\relax
\usepackage{diagbox}
\usepackage{tabularx}
\usepackage{amsmath}
\usepackage{amsfonts}
\makeatletter
\def\amsbb{\use@mathgroup \M@U \symAMSb}
\makeatother
\usepackage[bbgreekl]{mathbbol}
\newcommand{\ve}[1]{\ensuremath{\mbox{\boldmath$#1$}}}

\newcommand{\phm}{{\phantom-}}
\newcommand{\eqnlab}[1]{\label{eq:#1}}

\newcommand{\eqnref}[1]{(\ref{eq:#1})}

\newcommand{\obs}[1]{{#1}}

\begin{document}
\title{Inertial torque on a small spheroid in a stationary uniform flow}
\author{F. Jiang}
\address{SINTEF Ocean, \obs{NO-}7052 Trondheim, Norway}
\author{L. Zhao}
\address{\obs{AML,} Department of Engineering Mechanics, Tsinghua University, 100084 Beijing, China}
\author{H. I. Andersson}
\address{Department of Energy and Process Engineering, NTNU, NO-7491 Trondheim, Norway}
\author{K. Gustavsson}
\address{Department of Physics, Gothenburg University, \obs{Se-}41296 Gothenburg, Sweden}
\author{A. Pumir}
\address{Univ. Lyon, ENS de Lyon, Univ. Claude Bernard, CNRS, Laboratoire de Physique, F-69342,
Lyon, France}
\author{B. Mehlig}
\address{Department of Physics, Gothenburg University, 41296 Gothenburg, Sweden}

\begin{abstract}
How anisotropic particles rotate and orient in a flow  depends on the hydrodynamic torque they experience. Here we compute the torque acting on a small spheroid in a uniform flow by  numerically solving the Navier-Stokes equations. Particle shape is varied from oblate (aspect ratio $\lambda = 1/6$) to prolate ($\lambda = 6$), and we consider low and moderate particle Reynolds numbers (${\rm Re} \le 50$).  We demonstrate that the angular dependence of the torque, predicted theoretically for small particle Reynolds  numbers remains qualitatively correct for Reynolds numbers up to   ${\rm Re} \sim 10$. The amplitude of the torque, however, is smaller than the theoretical prediction, the more so as ${\rm Re}$  increases. For Re larger than $10$, the flow past oblate spheroids acquires a more complicated structure, resulting in systematic deviations from the theoretical predictions. Overall, our numerical results provide a justification of recent theories for the orientation statistics of ice-crystals settling in a turbulent flow.
\end{abstract}

\maketitle

\section{Introduction}
How does a spheroidal particle settle in a quiescent fluid?
When the settling velocity is small enough,
so that the fluid motion induced by the particle can be described by
the Stokes approximation \cite{Bre83,Kim:2005}, the
particle settles at an arbitrary
constant orientation equal to its initial orientation.
But since the initial
particle orientation is marginally stable, any small perturbation  must affect
the particle orientation. For example, for very small
particles, upon which thermal noise plays a significant role,
Brownian torques induce random orientation.
In addition, slight breaking of the fore-aft symmetry of the
particle \cite{Kha89,Can16,roy2019inertial} gives rise to a torque causing
the particle to settle at a steady angle determined by particle shape,
independent of its initial orientation.
These torques, induced either by thermal fluctuations or by
specific fore-aft asymmetry of the particle,
compete with the inertial torque arising from convective inertial corrections
to the Stokes approximation.
A  heavy particle settling steadily in a fluid experiences an undisturbed
uniform mean flow corresponding to the negative settling velocity. This mean
flow exerts a convective inertial torque on the particle. Its effect depends
upon the particle Reynolds number
 \begin{equation}
\label{eq:rep}
{\rm Re} = Ua_{\rm max}/\nu\,.
\end{equation}
Here $U$ is the settling speed of the particle, $\nu$ is the kinematic viscosity of the fluid, and $a_{\rm max}$ measures the maximal linear size of the
particle -- the half length of a rod or the radius of a disk.
For small ${\rm Re}$, the convective inertial torque turns the spheroid so that it settles with its broad side down.  \obs{Cox} \cite{Cox65} calculated the torque by perturbation theory in ${\rm Re}$, for nearly spherical particles in a uniform flow. A technically important point is that the convective-inertia  torque induced by the flow results from a singular perturbation of the Stokes equation, so
that straightforward perturbation theory in ${\rm Re}$ fails even at very small
values of ${\rm Re}$. Using asymptotic matching methods~\cite{Ben78}, Khayat \& Cox \cite{Kha89} obtained the convective-inertia torque in the slender-body limit, complementing the earlier results for nearly spherical particles. More recently, Dabade {\em et al.} \cite{Dab15} used the reciprocal theorem to calculate this torque for spheroids of arbitrary aspect ratio -- disks and rods -- to  linear order in ${\rm Re}$.

Several earlier numerical studies have been devoted to a determination of
the torque acting on spheroids in a uniform flow.
H\"olzer and Sommerfeld \cite{Holzer09} used a lattice-Boltzmann method (LBM) to compute the steady-flow torque on non-spherical particles of different shapes, amongst others for a prolate spheroid ($\lambda = 3/2$) at  different angles of inclination to the flow. Ouchene {\em et al.} \cite{Ouchene15,Ouchene16} used a commercial Navier-Stokes solver to resolve the flow field around prolate spheroids with aspect ratios $\lambda$ ranging from 5/4 to 32/1. 
Their more recent results for oblate spheroids are summarised in Ref.~\cite{Ouchene20}. Zastawny {\em et al.} \cite{Zastawny12} considered both prolate ($\lambda$ = 5/4 and 5/2) and oblate ($\lambda$ = 1/5) spheroids by means of an immersed boundary method and Sanjeevi {\em et al.} \cite{Sanjeevi18} used a LBM approach to compute the flow field around a prolate ($\lambda$ = 5/2) and an oblate ($\lambda$ = 2/5) spheroid at various angles of inclination. These earlier studies provide important insight for several spheroid
shapes. However, they give the torque only for
certain shapes, particle inclination to the flow, and particle
Reynolds number.  For slender fibres more is known. Shin {\em et al.} \cite{Shi05} performed numerical simulations, and their Fig.~5 shows that the Khayat \& Cox theory works well for slender fibers up to  Reynolds numbers of the order of $\sim 10$.
Finally, Zastawny {\em et al.} \cite{Zastawny12}, Ouchene {\em et al.} \cite{Ouchene16,Ouchene20}, Sanjeevi {\em et al.} \cite{Sanjeevi18}, and Fr\"ohlich {\em et al.}~\cite{Froehlich20}  proposed empirical parameterisations of the torque in the form of explicit functions of inclination angle and Reynolds number. We compare with some of these results later on.

\begin{figure}[b]
\centering
\begin{overpic}[width=12cm,clip]{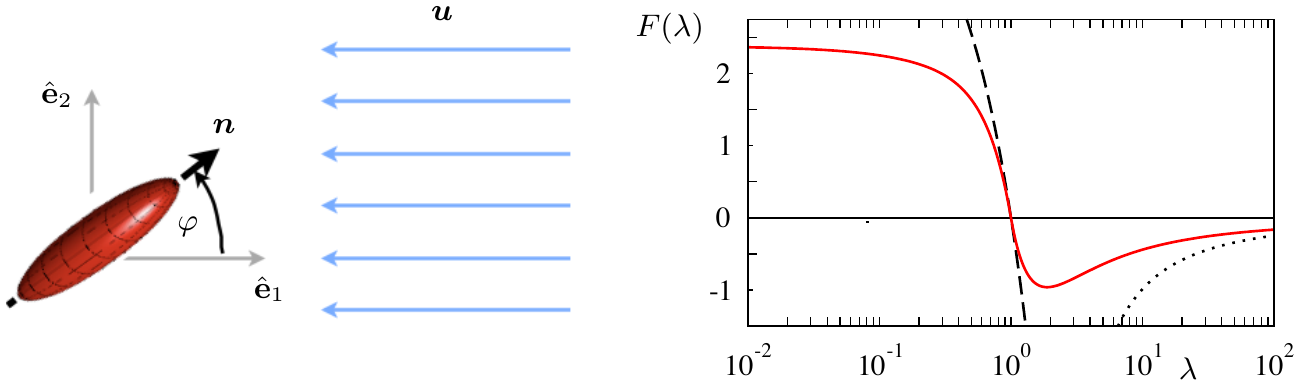}
\end{overpic}
\caption{\label{fig:schematic}  Left: prolate spheroid with symmetry axis $\ve n$  in a uniform flow with velocity $\ve u =     -U \hat{\bf e}_1$. The angle $\varphi$ between $\ve n$
and the $\hat{\bf e}_1$-axis is called tilt angle. Right: shape factor $F(\lambda)$ determining the torque $\tau_3$, Eq.~(\ref{eq:theory2}), red solid line. Also shown are the slender-body
asymptote (\ref{eq:slbl}), black dotted line, as well as the near-spherical asymptote (\ref{eq:nsph}), black dashed line.}
\end{figure}

Our goal is to validate the  small-Re-model \cite{Cox65,Kha89,Dab15}  for spheroids of different aspect ratios in a steady homogeneous flow, and to determine how the torque changes as the Reynolds number increases. To answer this question, we
solved numerically the Navier-Stokes equations past a spheroid at
rest  in a uniform and steady flow, as schematically
illustrated in Fig.~\ref{fig:schematic}(left)
at several values of the Reynolds number, ${\rm Re}$, and of the
particle shape (aspect ratio of the spheroid), in the case of
small platelets and of small columns.

\section{Method}
\label{sec:method}

Much of the literature on viscous and convective torques  on small non-spherical particles in a flow uses spheroids as model shapes because the resistance tensors that determine the motion of the particle in the fluid are known  \cite{Kim:2005}, and because fore-aft and rotational symmetry lead to a comparatively simple angular
dynamics.  In the following we consider spheroidal particles. Similarities and differences between the angular dynamics of spheroids and crystals with
discrete rotation and reflection symmetry were discussed by Fries {\em et al.} \cite{Fri17}.

We denote
the symmetry axis of the spheroidal particle by $\ve n$. The length of the symmetry axis is $2a_\parallel$, and the diameter of the spheroid is $2a_\perp$.  The aspect ratio of the spheroid  is
defined as $\lambda=a_\parallel/a_\perp$. Oblate particles (platelets) have $\lambda < 1$, while prolate particles (columns) have $\lambda >1$.
The Reynolds number defined in Eq.~(\ref{eq:rep}) is based upon $a_{\rm max}=\mbox{max}\{a_\parallel,a_\perp\}$.
We consider a small spheroidal particle at a fixed position in a steady homogeneous flow with velocity $\ve u$.
For a prolate spheroid, the setup is shown in Fig.~\ref{fig:schematic}(a). The tilt angle $\varphi$ is defined as the angle between the particle-symmetry vector $\ve n$ and $-\ve u$,
for prolate as well as for oblate spheroids.  For fore-aft symmetric particles, it is sufficient to consider angles $\varphi$ in the interval $[0,\tfrac{\pi}{2}]$.

The  hydrodynamic torque with respect to the centre of the particle reads
\begin{equation}
\ve  \tau=  \int_{\mathscr{S}}\!\! \ve r\wedge(\bbsigma{\rm d}\ve s) \,.
\label{eq:torque}
\end{equation}
Here  $\sigma_{mn}  = -p\delta_{mn} + 2 \mu S_{mn}$  are the elements of the stress tensor $\bbsigma$ with  pressure $p$,  $S_{ij}$ are the elements of the strain-rate tensor of the disturbance flow, and $\mu=\rho_{\rm f} \nu$ is the dynamic viscosity
with fluid-mass density $\rho_{\rm f}$. The integral in Eq.~(\ref{eq:torque}) is over the particle surface $\mathscr{S}$, $\ve r$ is the displacement vector from the particle centre to a point on the particle surface, and
${\rm d}\ve s$ is the outward surface normal at this point.
We computed  the torque by numerically solving
the full three-dimensional Navier-Stokes equations for incompressible flow, using the solver MGLET  \cite{manhart2001mglet}.
This  code was recently used to document the computational challenges of calculating forces and torques upon rods
in  uniform flows \cite{Helge}. The method is briefly described in appendix \ref{app:A}. 
To quantify the convective-inertial effect for particles of different sizes and shapes, we
fixed the Reynolds number (\ref{eq:rep})
as we varied particle shape. 
Both viscous stress and pressure contribute to the torque in general,
but at small Reynolds numbers the 
the viscous stresses dominate.

Before discussing the small-Re theory \cite{Cox65,Kha89,Dab15}, consider first how symmetries constrain the torque
(an equivalent derivation was first given in Ref.~\cite{subramanian2005}).
The torque $\ve\tau$ is a pseudovector, it transforms as a vector under
rotations, and with a sign change under reflections. The dependence of the torque on \obs{the vectors $\ve n$ (particle orientation) and $\hat{\ve u}$ (flow direction)}
 is obtained as a linear combination of products of the components \obs{of} $\ve n$ and $\hat{\ve u}$, and the primitive invariant tensors under proper rotations, 
the Kronecker delta  $\delta_{ij}$ and \obs{the Levi-Civita symbol} $\varepsilon_{ijk}$.
The torque depends only on the relative orientation between the spheroid and the flow
direction. Therefore, and since a spheroid is fore-aft symmetric, the torque must be invariant under  $\hat{\ve u}\to -\hat{\ve u}$.
The only pseudovector satisfying these symmetries is
\begin{equation}
\label{eq:tau_general}
\ve\tau=g(U,\ve n\cdot\hat{\ve u})(\ve n\wedge\hat{\ve u})\,,
\end{equation}
where $U = |\ve u|$, and $g$ is an odd function \obs{of} $\ve n\cdot\hat{\ve u}$ .
To leading order in Re, the strain $\bbsigma$ balances the nonlinear
term of the Navier-Stokes equations, and as a result, the
magnitude of the hydrodynamic torque on a small spheroid in a uniform flow $\ve u$  is proportional to $U^2$, to leading order in  ${\rm Re} = U a_{\rm max}/\nu$. Therefore 
$g(U,\ve n\cdot\hat{\ve u}) \propto U^2(\ve n\cdot {\hat{\ve u}})$, so that \cite{subramanian2005}$\ve \tau^{(2)}\propto U^2 (\ve n\cdot {\hat{\ve u}})(\ve n\wedge {\hat{\ve u}})$. The superscript $(2)$ emphasises that $\ve \tau^{(2)}$ refers to the leading-order torque, quadratic in $U$.
Dimensional analysis shows that the dimensional factors must be of the form  $F(\lambda){\rho_{\rm f}}   {U^2a_{\rm max}^3}$,
where $F(\lambda)$ is a shape factor that depends on the particle aspect ratio but not its size. In short,
\begin{align}
&\ve\tau^{(2)}= F(\lambda){\rho_{\rm f}}   {U^2a_{\rm max}^3}\,
 (\ve n\cdot {\hat{\ve u}})(\ve n\wedge {\hat{\ve u}})\,
\eqnlab{torque_fluid_inertia}
\end{align}
for small ${\rm Re}$. 

For general values of ${\rm Re}$, Eq.~(\ref{eq:tau_general}) implies that
the  torque  resulting from the steady flow past the object  must be 
perpendicular to $\ve n$ and $\hat{\ve u}$, and that it is proportional to $\sin\varphi$ times a function odd in $\cos\varphi$.
This means that the torque must vanish at $\varphi=0$ and $\varphi=\pi/2$.
The leading-order torque $\ve\tau^{(2)}$, given by Eq.\eqnref{torque_fluid_inertia},
is proportional to $\cos\varphi\sin\varphi$ and thus symmetric around $\varphi=\pi/4$. 
This angular dependence provides a good qualitative description of earlier numerical results for
the angular dependence of the torque, see e.g. Ref.~\cite{Sanjeevi18}, but
 Eq.~(\ref{eq:tau_general}) implies that higher-order corrections to the torque 
contain terms with higher (odd) powers of $\cos\varphi$, breaking the symmetry
around $\varphi = \pi/4$ when ${\rm Re}$ becomes large enough.

The shape factor $F(\lambda)$ computed in Ref.~\cite{Dab15} is shown in Fig.~\ref{fig:schematic}(b). Also shown
is the slender-body limit
\begin{equation}
\label{eq:slbl}
F(\lambda) \sim -5\pi/[3(\log\lambda)^2]\,,
\end{equation}
as well as the near-spherical expansion~\cite{Dab15}
\begin{equation}
\label{eq:nsph}
F(\lambda) \sim \mp 811 \pi\varepsilon/560
\end{equation}
 for small eccentricity $\varepsilon$.
Here the eccentricity parameter is defined by $\lambda = 1+\varepsilon$
for prolate particles, and $\lambda = (1-\varepsilon)^{-1}$ for oblate particles.
The slender-body limit (\ref{eq:slbl}) agrees with that derived earlier by Khayat and Cox \cite{Kha89}, but
Eq.~(\ref{eq:nsph})  differs slightly from the result of  Cox \cite{Cox65} for nearly spherical particles, as mentioned
in Ref.~\cite{Dab15}.

In the following we assume without loss of generality that the uniform flow points along the negative $\hat {\bf e}_1$-axis,
$\ve u = -U \hat {\bf e}_1$, and that the symmetry vector $\ve n$ lies in the  $\hat {\bf e}_1$-$\hat {\bf e}_2$-plane  (Fig.~\ref{fig:schematic}). Then the torque aligns with the  $\hat {\bf e}_3$-axis, $\ve \tau = \tau_3  \hat {\bf e}_3$, where $\hat {\bf e}_3=\hat {\bf e}_1\wedge \hat {\bf e}_2$. In this case Eq.~\eqnref{torque_fluid_inertia} implies that the torque depends on the tilt angle $\varphi$  as
\begin{equation}
\label{eq:theory}
\tau_3^{(2)}=- \tfrac{1}{2}F(\lambda){\rho_{\rm f}}   {U^2a_{\rm max}^3}\,
\sin2\varphi\,.
\end{equation}
As mentioned above,
the torque $\tau_3$  vanishes for $\varphi = 0$,
corresponding to $\hat{\bf n}$ parallel to  $\hat{\ve u}$, and for
$\varphi = \pi/2$, when $\hat{\bf n} $ and $\hat{\ve u}$ are perpendicular to
each other.  The sign of $F(\lambda)$ [Fig.~\ref{fig:schematic} (right)] implies that   $\varphi = \tfrac{\pi}{2}$ is
stable for prolate particles (rods), whereas $\varphi = 0$
is stable for oblate particles (disks).

\section{Numerical results}
\label{sec:results}
\begin{figure}[t]\centering
\begin{overpic}[width=14.cm,clip]{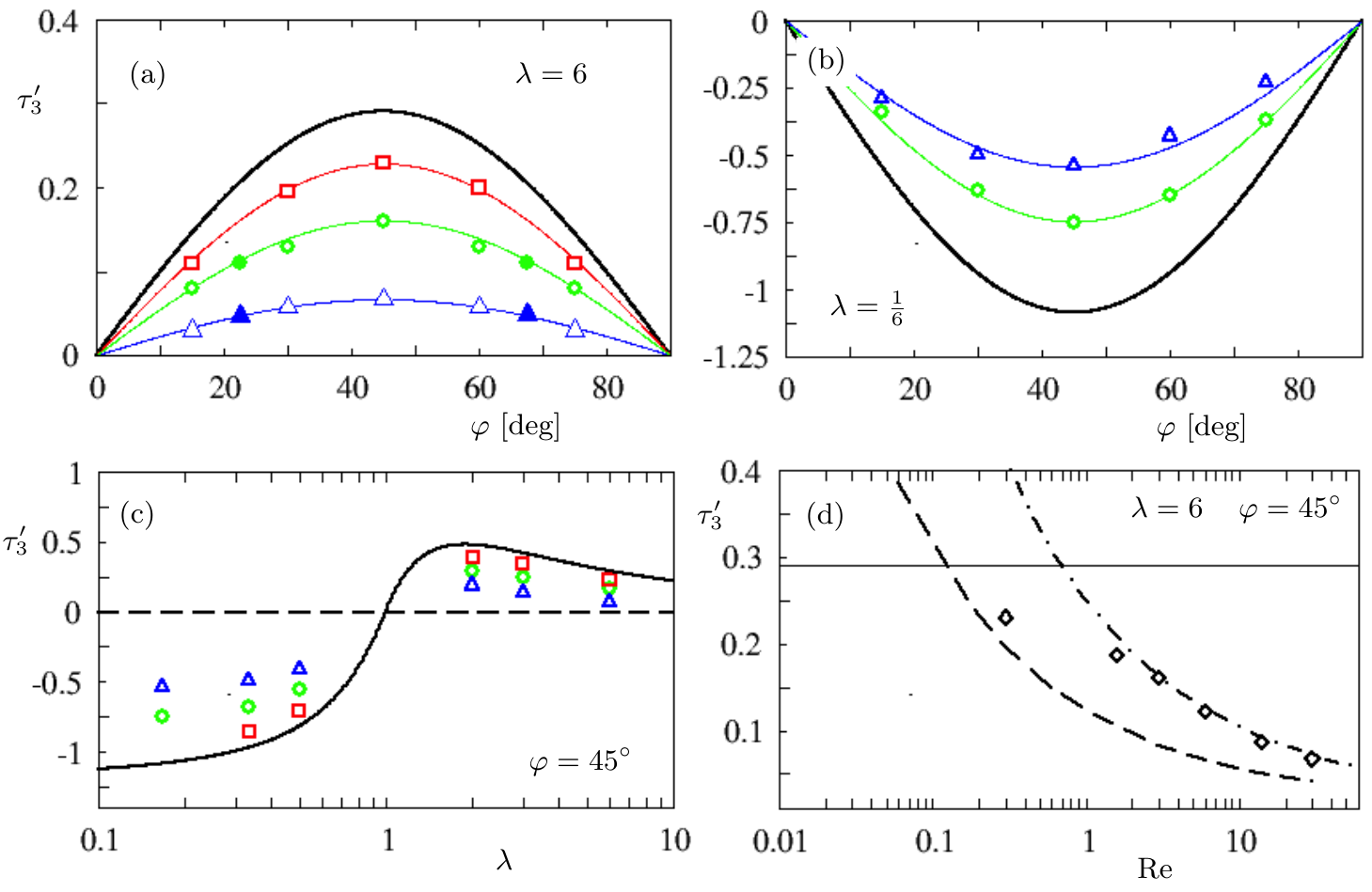}
\end{overpic}
\caption{\label{fig:torque} (a) Dimensionless torque $\tau_3'$ [Eq.~(\ref{eq:dimlesstau})]
upon a prolate spheroid in a uniform flow, as a function of the angle of inclination. Results  for  $\lambda=6$ and ${\rm Re} = 0.3$ (red, $\Box$), ${\rm Re}=3$ (green, $\circ$),
and ${\rm Re} = 30$ (blue, $\triangle$). Filled symbols correspond to data from Ref.~\cite{Helge}.  Theory (\ref{eq:theory2}) is shown as a solid black line. Coloured lines are fits of
the angular dependence in Eq.~(\ref{eq:theory2}),  proportional to $\sin2\varphi$. (b) Same for an oblate spheroid with $\lambda = 1/6$. (c) Maximal torque $\tau_3'$ (evaluated at
$\varphi= 45^\circ$) as a function of aspect ratio $\lambda$, for different particle Reynolds numbers. Symbols show simulation results,
the black solid line is $\tfrac{1}{2}F(\lambda)$. (d) Dependence of  $\tau_3'$ upon Reynolds number  for  $\varphi = 45^\circ$, $\lambda=6$ (black, $\Diamond$), compared with the small-Re theory (\ref{eq:theory2}), solid line,  with the parameterisation of Ouchene {\em et al.} \cite{Ouchene16}, dashed line, and  with the parameterisation of Fr\"ohlich {\em et al.} ~\cite{Froehlich20}, dash-dotted line. }

\end{figure}
We de-dimensionalise the torque as
\begin{equation}
\label{eq:dimlesstau}
\tau_3' = \frac{\tau_3}{\rho_{\rm f} U^2 a_{\rm max}^3}\,.
\end{equation}
Fig.~\ref{fig:torque} shows our simulation results for the dimensionless torque
 for prolate and oblate spheroids (appendix \ref{app:B}), compared with the small-Re theory (\ref{eq:theory}) which reads in dimensionless form:
 \begin{equation}
 \label{eq:theory2}
 \tau_3' =- \tfrac{1}{2} F(\lambda) \sin2\varphi\,.
 \end{equation}
This theory is shown as a thick solid line. Panel (a) contains the results for a prolate spheroid with $\lambda =6$  as a function of tilt angle, for different particle Reynolds numbers (symbols). Filled symbols correspond to data from Table~5 in Ref.~\cite{Helge}.  Thin solid lines are fits to the theoretically predicted angular dependence, proportional to $\sin 2\varphi$. We see that the numerical results for the smallest Reynolds number, ${\rm Re}=0.3$, agree quite well with the theory, the deviation is about 20\%. For larger Reynolds numbers the deviations are larger, but the angular dependence is still accurately predicted by the small-Re theory, only the amplitude becomes smaller.

Panel (b) shows results for an oblate spheroid with aspect ratio $\lambda =\tfrac{1}{6}$. The results are qualitatively similar to those obtained for the prolate particle, but there are two important
differences. First, we have no data points for ${\rm Re}=0.3$. The smallest Reynolds-number simulations are very costly
because one must use a large domain size at the same time as a small spatial mesh \cite{Helge}. This is particularly challenging
for disks  because a finer mesh is needed to resolve the flow in the vicinity of the strongly curved periphery of flat disks.  The second difference is that for the disk, the $\varphi$-dependence develops an asymmetry  around
$\varphi=45^\circ$ at larger values of Re.

Panel (c) shows the torque at $\varphi=45^\circ$ as a function of particle aspect ratio in comparison with Eq.~(\ref{eq:theory2}). We infer that the theory describes
the shape dependence of the inertial torque well, quantitatively at ${\rm Re}=0.3$, and qualitatively at  the larger Re.

Panel (d)  shows our numerical result for the torque for $\lambda=6$ and $\varphi=45^\circ$ as a function of Reynolds number, compared with the parameterisations of Ouchene {\em et al.} \cite{Ouchene16} and Fr\"ohlich  {\em et al.}~\cite{Froehlich20}, and with the small-Re limit (\ref{eq:theory2}).
At larger Re, we observe excellent agreement with the parameterisation of Fr\"ohlich  {\em et al.}~\cite{Froehlich20}.
Agreement with the parameterisation of Ouchene {\em et al.} is not as good. 
At any rate, all parameterisations  appear to be inconsistent 
with the theory (\ref{eq:theory2}) at small Re, they predict a much larger torque in this limit.

 \begin{figure}[t]
\centering
 \begin{overpic}[width=8cm]{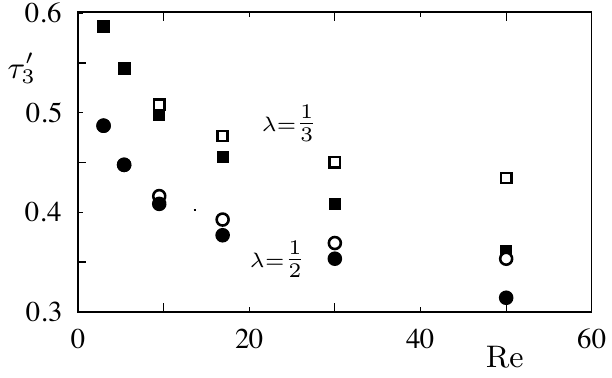}
  \end{overpic}
\caption{\label{fig:torque_disk_large_Re}
 Torque on a disk as a function of Reynolds number. Results for spheroids
of two different shapes are shown: $\lambda = 1/2$ (circles),  $\lambda = 1/3$ (squares).
Empty symbols correspond to tilt angle $\varphi = 30$, and
full symbols to $\varphi = 60$. At ${\rm Re} = 3$ and $5$,
empty and full symbols lie on top of each other.
}
\end{figure}

Fig.~\ref{fig:torque_disk_large_Re}
quantifies the asymmetry of the $\varphi$-dependence of the torque around $\varphi=45^\circ$ that develops for disks at larger Reynolds numbers. We note that our initial configuration is symmetric w.r.t. reflection in the $x$-$y$ plane [Fig.~\ref{fig:schematic}(left)]. We have checked that the flow remains 
symmetric and 
steady for all simulations described in this article, for Reynolds numbers up to ${\rm Re}=50$.
\begin{figure}[t]
 \centering
 \begin{overpic}[width=12cm]{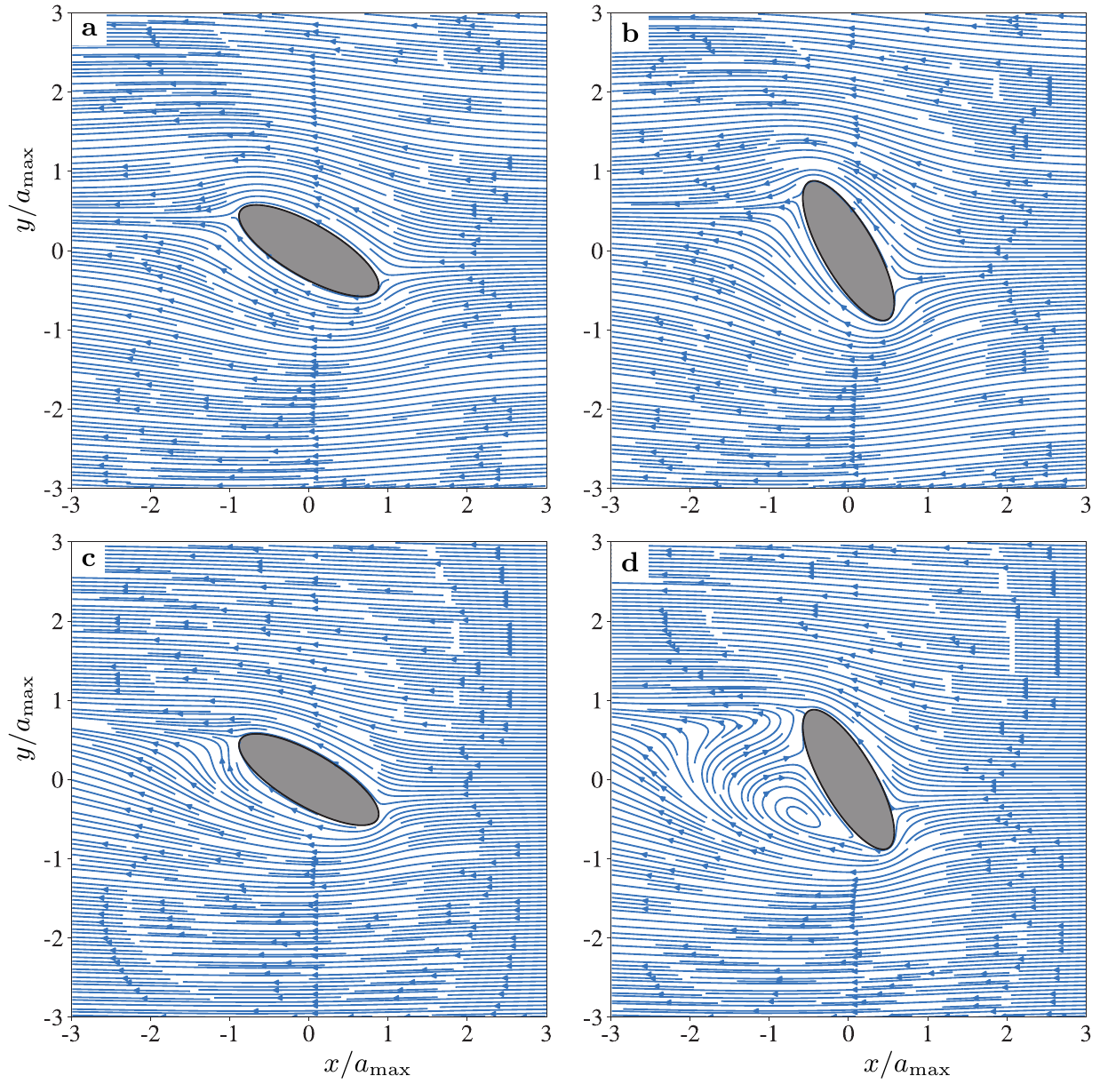}
\end{overpic}    \caption{\label{fig:disturbance} Streamlines of the flow  around a disk
in the $x$-$y$ plane at $z = 0$, for ${\rm Re}= 3$ and $\lambda = 1/3$. The tilt angle is (a) $\varphi=30^\circ$ and (b) $\varphi=60^\circ$;
(c) and (d) show the same but for ${\rm Re}= 30$.  }
 \end{figure}

 \section{Discussion}
 \label{sec:discussion}
Our results shown in Fig.~\ref{fig:torque} demonstrate 
reasonable agreement between the numerical-simulation results at small Reynolds
numbers and the theory \eqnref{torque_fluid_inertia}. The shape dependence remains  qualitatively correct  for
the largest Reynolds numbers in Fig.~\ref{fig:torque}, Re$=30$.  But in general the torque is smaller than the small-Re theory \eqnref{torque_fluid_inertia} predicts. For example,
Fig.~\ref{fig:torque}(b) shows that the maximal ${\rm Re}\!=\!30$-torque on a disk is smaller than the small-Re prediction by about a factor of two. As mentioned above and in Ref.~\cite{Dab15}, there are slightly different predictions \cite{Dab15,Cox65} for $F(\lambda)$ for nearly spherical particles. 
Our numerical simulations are not accurate enough to determine which of the two predictions is correct, as the difference is much smaller than that
documented in Fig.~\ref{fig:torque}(a,b). 

Turning to the comparison between our numerical results, theory, and
the parameterisations by Ouchene {\em et al.}  \cite{Ouchene16}
and Fr\"ohlich  {\em et al.}~\cite{Froehlich20},  the excellent agreement between
our simulation results and the  parameterisation of Fr\"ohlich  {\em et al.} at large Re [Fig.~\ref{fig:torque}(d)] indicates that  their and our numerical simulations are consistent in this range.
As mentioned above, the parameterisation of Ouchene {\em et al.}  does not agree as well. 
A possible reason may be that  the mesh in the simulations of Ouchene {\em et al.}~\cite{Ouchene16}
 was not fine enough. Their recent simulation results for oblate spheroids \cite{Ouchene20} were obtained with a finer mesh, and they agree well with the results of our simulations (not shown).

At small Re, both parameterisations fail.
This is expected because they are
derived for and intended to work for larger values of Re. 
At small Re, by contrast, there are substantial deviations. In particular, the parameterisations do not appear to converge to the correct limit (\ref{eq:theory2}) as Re tends to zero. 
Our \obs{numerical} simulation results  agree somewhat better  with the theory at small Re. At the smallest Reynolds number we could simulate, Re$=0.3$, the relative error between the simulation results and theory is about 25\%.  Our simulation results exhibit a qualitative change in the Re-dependence for Re of order unity and smaller [panel (d)], likely due to the fact that the balance of terms responsible for the torque changes. Whereas pressure dominates for ${\rm Re} >1$, the viscous contribution becomes more important for small Reynolds numbers. We note that the ratio between the pressure and the viscous contributions decreases from $\approx 2.54$ at ${\rm Re} = 30$ down to $\approx 1.37$ at ${\rm Re} = 0.3$ (not shown).

The small-Re theory for the torque exhibits a symmetry around $\varphi = 45^\circ $. 
For prolate particles our numerical simulations exhibit this symmetry quite accurately even
at the largest values of Re we simulated, and this is consistent
with the results of other numerical studies \cite{Holzer09,Ouchene15,Zastawny12,Jiang14}.

For disks, by contrast, this symmetry is clearly broken already at ${\rm Re}=10$, and this may imply that the maximum of the torque is not
precisely at $\varphi=45^\circ$. The asymmetry increases as the Reynolds number increases, and for $Re=50$, the relative
asymmetry is about $17\%$ for $\lambda = \obs{1/3}$.
Recall that the symmetry with respect to $\varphi = \pi/4$ holds at the lowest order in perturbation theory. As explained in Section~\ref{sec:method}, higher-order contributions are expected to break this symmetry. To get a more physical understanding of the mechanisms involved, we visualised the fluid-velocity field around a disk with aspect ratio $\lambda=1/3$ at $\varphi =30^\circ$ and $60^\circ$ in Fig.~\ref{fig:disturbance}.  We observe that the streamlines closely follow the surface of the spheroid in panels (a) and (b). This reflects that flow remains attached to the surface of this oblate spheroid at small Reynolds numbers. At ${\rm Re}=30$, by contrast, the flow separates as the oblate spheroid meets the flow with its broad side, resulting in quite different flow patterns for $\varphi =30^\circ$ and $60^\circ$. This certainly contributes to the asymmetry of the torque.
 It is likely that this asymmetry in the $\varphi$-dependence is a precursor of a bifurcation, as the Reynolds number increases.
Indeed, experiments show that there is a transition for a disk: it settles with its broad side down at small Re, but  exhibits other kinds of periodic or chaotic lateral and angular dynamics at larger Re,  due to interactions between the disk and the induced vortex street. A bifurcation to periodic angular dynamics happens at ${\rm Re}\sim 100$
\cite{willmarth1964steady,field1997chaotic}. 

In summary, the symmetry breaking -- clearly visible for ${\rm Re}
\gg 10$ for disks -- does not  develop for rods  in the Re-range we considered.
This difference must reflect higher-order 
corrections in a formal expansion of the torque in 
${\rm Re}$. A more physical understanding can be obtained by noting that flow detachment contributes to the torque asymmetry for disks (Fig.~\ref{fig:disturbance}). For rods, by contrast, their limited extent in the spanwise
direction impedes detachment at large ${\rm Re}$. As a consequence, the flow
perturbations for rods do not reveal as high a degree of asymmetry under
reflection of $\varphi$ with respect to $\pi/4$, compared with disks.
Note that the reflection symmetry with respect to the plane defined by 
the direction of the velocity vector and the the particle axis breaks
at much higher values of Re. Also, the flow becomes unsteady at very large Reynolds numbers. For example, the flow around a spheroid with $\lambda=6$ and tilt
angle $\varphi=45^\circ$ becomes unsteady  at about ${\rm Re}=\obs{1}000$  \cite{jiang2015transitional}.

The simulations described above were motivated in part by recent studies of the angular dynamics of ice platelets settling
in turbulent clouds  \cite{Kle95,Bre04,Kramel,Men17,Lop17,Gus19,Sha19,Gus20}.
A small platelet settling in a turbulent flow experiences a mean flow equal in magnitude to its settling speed, in addition to turbulent fluctuations which render the flow non-uniform
and unsteady. The standard model for ice-crystal dynamics in turbulence  \cite{Kle95,Bre04,Kramel,Men17,Lop17,Gus19,Sha19,Gus20}
assumes that the fluid torque on a settling crystal can be approximated by the superposition of the Jeffery torque due to the turbulent fluid-velocity gradients, and the small-Re expression for the convective inertial torque, Eq.~(\ref{eq:theory}). These two contributions compete, in that the convective inertial torque tends to align the particles, while the Jeffery torque tends to randomise their orientations, and this model was shown to qualitatively describe 
the results of experiments measuring the angular dynamics of small  rods settling in a cellular flow \cite{Lop17}.
The extent to which turbulence destroys alignment of settling
particles has important consequences in the atmospheric sciences, where reflection of polarised light reveal small orientation fluctuations of small ice crystals \cite{Pru78,chen1994theoretical} settling in turbulent clouds~\cite{Bre04}.
Therefore it is important to validate the assumptions underlying the model \cite{Kle95,Bre04,Kramel,Men17,Lop17,Gus19,Sha19,Gus20}.

First, 
shear-induced contributions to the inertial torque  \cite{einarsson2015a} are neglected. 
This approximation is justified for particles 
smaller than the Kolmogorov length $\eta_{\rm K}$
characterising the size of the smallest turbulent eddies \cite{Fri97}, since
 ${\rm Re}_s \sim (a/\eta_{\rm K})^2$ \cite{Candelier2016}, 
and the shear-induced inertial torque is negligible compared to inertial corrections (\ref{eq:theory}) when ${\rm Re} \gg \sqrt{{\rm Re}_s}$ \cite{Candelier2018}. 
Second, it is appropriate to use the steady theory (\ref{eq:theory}) if the time scale at which the
slip velocity changes is much smaller than the viscous time $a_{\rm max}^2/\nu$. This condition is well satisified for 
particle sizes of the order of the Kolmogorov length or smaller \cite{Gus20}. In the experiments of Lopez and Guazzelli \cite{Lop17},
this condition was marginally satisified \cite{Gus20}, nevertheless the model predictions agree qualitatively with the measurements.
Third the model assumes that the shape and angular dependence of the convective inertia torque due to the mean flow is given by
\eqnref{theory}. The results summarised here show that the expression for
the torque works qualitatively quite well 
for the regimes corresponding to crystals in clouds, with Reynolds
number up to $30$, see Table S1 in \cite{Gus20}. The reduction in the magnitude
of the torque seen in Fig.~\ref{fig:torque}a-c are not expected to lead to 
any significant qualitative changes of the model predictions.

\section{Conclusions}
We performed numerical simulations determining the hydrodynamic torque on oblate and prolate spheroids that settle steadily in a quiescent fluid. 
 We compared the numerical results with low-Re theory for the hydrodynamic torque, Eq.~(\ref{eq:theory}), and found  quantitative agreement for the smallest Reynolds numbers [Fig.~\ref{fig:torque}(c))]. Deviations at larger Reynolds numbers depend on particle shape. For prolate particles we found that the tilt-angle dependence remains $\sin2\varphi$ as predicted by the theory  for Re up to the largest Reynolds number we have simulated. This is consistent with the earlier numerical results of  Jiang, Gallardo \& Andersson \cite{Jiang14}.

For disks, by contrast, we found that this symmetry is broken 
 already at ${\rm Re}=10$. We attribute this symmetry breaking to the fact that the flow detaches from the disk when it faces the flow with its broad side. 
For both prolate and oblate particles we found that the torque amplitude is smaller than the theoretical
 prediction. The difference is small when Re is very small, and it grows as Re grows. 
 
 We also compared our numerical results with the parameterisations of the hydrodynamic torque proposed
 by Ouchene {\em et al.} \cite{Ouchene16}  and Fr\"ohlich  {\em et al.}~\cite{Froehlich20},  which model  not only the dependence of the torque on Re and on the tilt angle, but also upon the aspect ratio. 
 We found that the parameterisations both fail to reproduce the theory in the small-Re limit.

Here we studied the small-Re limit of the problem.
Experiments at large Re (${\rm Re}\sim 1000$)  compare the  trajectories and velocities of platelets settling in a quiescent fluid, with those settling in a turbulent background flow
\cite{esteban2020disks}. The authors find that the background turbulence has a significant effect upon the settling dynamics.
This is expected because the fluid-velocity gradients give rise to Jeffery torques, as mentioned above. 

To validate the model used in Refs.~\cite{Kle95,Kramel,Men17,Lop17,Gus19,Sha19,Gus20}, it would be of interest to conduct experiments at smaller Reynolds numbers, so that one can compare and contrast with the predictions of Refs.~\cite{Gus19,Gus20}, for example. We intend to run fully resolved simulations of particles settling in turbulence in order to justify and refine the model. But this remains a challenge for the future.

\acknowledgments
KG and BM were supported by the grant {\em Bottlenecks for particle growth in turbulent aerosols} from the Knut and Alice Wallenberg Foundation, Dnr. KAW 2014.0048, and in part by VR grant no. 2017-3865 and Formas grant no. 2014-585.  AP  acknowledges support from the IDEXLYON project (Contract ANR-16-IDEX-0005) under University of Lyon auspices.  
BM and LZ were supported by a collaboration grant from the joint China-Sweden mobility programme \obs{[National Natural Science Foundation of China (NSFC)-Swedish Foundation for International Cooperation in Research and Higher Education (STINT)]},  grant numbers 11911530141 (NSFC) and CH2018-7737 (STINT).
LZ acknowledges support from the Natural Science Foundation of China (grant no. 91752205). 
FJ acknowledges support from SINTEF Ocean via. internal funding in numerical hydrodynamic group.

\mbox{}

\newpage
\appendix
\section{Description of simulations}
\label{app:A}

MGLET is a finite-volume code that directly solves the full time-dependent three-dimensional Navier-Stokes equations for incompressible fluids. The computational domain is discretised on a multi-level staggered Cartesian mesh with cubic grid cells. A third-order explicit low-storage Runge-Kutta scheme \cite{Williamson80} is used for the time dependence. Stone's strongly implicit procedure \cite{Stone68} is applied for pressure correction in each time step. To represent the curved particle surface in the Cartesian mesh, MGLET uses a direct-forcing immersed-boundary method (IBM), representing no-slip and impermeable boundary conditions at the particle surface. The code has been extensively validated for different flows over a wide range of Reynolds numbers, among which Refs.~\cite{Jiang14, Helge} are in the low-Re regime and therefore most relevant to the present study.

Within the immersed boundary method, separate force components in the three Cartesian directions are obtained by summing up unbalanced momentum fluxes at the intersected mesh cell faces. The pressure is explicitly solved during the simulation and allows a direct integration over the surface of the spheroidal body. Viscous forces are therefore obtained by subtracting the pressure force contribution from the total force. The fact that all forces are directly obtained in the three Cartesian directions also ensures a straightforward torque calculation. The resulting torque is consistent with Eq. (\ref{eq:torque}).

All simulations in the present study used the largest practically possible computational domain ($34 a_{\rm max} \times 34 a_{\rm max} \times 34 a_{\rm max}$).
 The minimum grid cell size was $0.0033\,a_{\rm max}$. The relatively fine mesh and large computational domain lead to a large mesh size
 (of the order of \obs{$10^7$} grid cells), and the explicit time-evolution scheme requires a very small time step when
 the Reynolds number is small. These challenges are discussed in Ref.~\cite{Helge}.

Here we define the Reynolds number using $a_{\rm max}$, Eq.~(\ref{eq:rep}). The authors of Ref.~\cite{Helge}  define the Reynolds number
(${\rm Re}_D$ in their notation) in terms of the short-axis length $D=2a_{\rm min} $,
where $a_{\rm min} = \mbox{min}\{a_\parallel,a_\perp\}$.
To determine the effect of particle shape, we varied the aspect ratio $\lambda$ while keeping $a_{\rm min}$ constant.  It follows
that ${\rm Re}_D$ and ${\rm Re}$ defined in Eq.~(\ref{eq:rep})  are related as:
\begin{equation}
\label{eq:comparison}
{\rm Re} = \tfrac{1}{2} {\rm Re}_D\left\{ \begin{array}{ll}
\lambda \quad& \mbox{for $\lambda > 1$,}\\
\lambda^{-1} \quad &\mbox{for $\lambda < 1$.}
\end{array}\right .
\end{equation}
For the aspect ratio $\lambda =6$ studied
in Ref.~\cite{Helge} we have ${\rm Re} = \tfrac{\lambda}{2} {\rm Re}_D = 3 Re_{\rm D}$.
The authors of Ref.~\cite{Helge}  also defined a second Reynolds number, ${\rm Re}_p$ in their notation, in terms of the sphere-equivalent diameter $d_0 = 2a_0$.
Since the volume of the spheroid is $\tfrac{4\pi}{3} a_\parallel a_\perp^2$, we have that $a_0 = (a_\parallel a_\perp^2)^{1/3} = \lambda^{1/3} a_\perp = \lambda^{-2/3} a_\parallel$.
The authors of Ref.~\cite{Helge} de-dimensionalise the torque by dividing by $\tfrac{1}{2} \rho_{\rm f} U^2 \tfrac{\pi}{8} d_0^3$. To compare with their results for $\lambda = 6$
we use
\begin{equation}
d_0 = 2a_0 = 2 a_{\rm max}
\left\{ \begin{array}{ll}
\lambda^{-2/3}  \quad& \mbox{for $\lambda > 1$,}\\
\lambda^{1/3}  \quad &\mbox{for $\lambda < 1$.}
\end{array}\right .
\end{equation}
The ratio of normalisation factors is
\begin{equation}
\frac{\tfrac{1}{2}\rho_{\rm f}U^2 \tfrac{\pi}{8} d_0^3}{\rho_{\rm f}U^2a_{\rm max}^3}
=\frac{\pi}{2} \left\{ \begin{array}{ll}
\lambda^{-2}  \quad& \mbox{for $\lambda > 1$,}\\
\lambda  \quad &\mbox{for $\lambda < 1$.}
\end{array}\right .
\end{equation}
\newpage
\section{Summary of simulation results}
\label{app:B}
\begin{table}[h]
\caption{\label{tab:torque} Numerical results (MGLET)  for torque $\tau_3' = \tau_3/( \rho_{\rm f} U^2 a_{\rm max}^3)$ upon a spheroid in a uniform flow,
as a function of tilt angle $\varphi$, Reynolds number Re, and particle aspect ratio $\lambda$.
 }

${\rm Re}=0.3$\\[1mm]
\begin{tabular}{llllllll}\hline\hline
\diagbox{\small $\lambda$}{\small $\varphi$ [deg] }   & 15   & 30     & 45   & 60   & 75  \\\hline
6  & \phm 0.112 & \phm 0.196 & \phm 0.226 & \phm 0.199 & \phm 0.114 \\
3   &  & &  \phm 0.340&  &  & &   \\
2   &  & &  \phm 0.393&  &  &  &  \\
$\tfrac{1}{2}$  &  & & -0.707  &  &  & &   \\
$\tfrac{1}{3}$   &  & & -0.853 &  &  & &   \\
$\tfrac{1}{6}$   &  & &  &  &  & &   \\\hline\hline
\end{tabular}

${\rm Re}=3$\\[1mm]
\begin{tabular}{lllllllll}\hline\hline
\diagbox{\small $\lambda$}{\small $\varphi$ [deg]}  & 15   & 30     & 45   & 60     &  75  \\\hline
6  & \phm 0.076  & \phm 0.133 & \phm 0.159 & \phm 0.135 & \phm 0.078 &\\
3  & \phm 0.120  & \phm 0.211 &\phm  0.244 & \phm 0.213 & \phm 0.122 &  \\
2  & \phm 0.145 & \phm 0.251 &  \phm 0.291 & \phm 0.255 & \phm  0.145 &  \\
$\tfrac{1}{2}$  & -0.283 &-0.487& -0.558&  -0.487      & -0.275  \\
$\tfrac{1}{3}$ & -0.340  & -0.586& -0.681  &-0.586   &  -0.335 \\
$\tfrac{1}{6}$ & -0.340 &  -0.628 & -0.746 & -0.649 &  -0.369\\\hline\hline
\end{tabular}

${\rm Re}=30$\\[1mm]
\begin{tabular}{lllllllll}\hline\hline
\diagbox{\small $\lambda$}{\small $\varphi$ [deg]}   & 15   & 30     & 45   & 60     &  75 \\\hline
6  & \phm 0.033  & \phm 0.057 & \phm 0.065 & \phm 0.057&  \phm 0.033  \\
3  & \phm 0.068  & \phm 0.117 & \phm 0.136 & \phm 0.119&  \phm 0.068  \\
2 & \phm 0.094  & \phm 0.161 &\phm -0.185 & \phm 0.161 & \phm 0.090  \\
$\tfrac{1}{2}$  & -0.220  & -0.369 & -0.416   & -0.353  & -0.196  \\
$\tfrac{1}{3}$  & -0.267  & -0.450  & -0.497   &   -0.408  & -0.225 \\
$\tfrac{1}{6}$  & -0.293   & -0.503 & -0.547 & -0.432  &-0.233 \\\hline\hline
\end{tabular}
\end{table}

\end{document}